# Origin and Evolution of Jupiter's Trojan Asteroids

William F. Bottke (1), Raphael Marschall (2), David Nesvorný (1), David Vokrouhlický (3)

(1) Southwest Research Institute, Boulder, CO, USA, bottke@boulder.swri.edu
(2) CNRS, Observatoire de la Cote d'Azur, Laboratoire J.-L. Lagrange, CS 34229, 06304 Nice Cedex 4, France
(3) Institute of Astronomy, Charles University, V Holešovičkách 2, CZ-18000, Prague 8, Czech Republic

Prepared for **Space Science Reviews**

October 19, 2023

## Abstract

The origin of the Jupiter Trojan asteroids has long been a mystery. Dynamically, the population, which is considerably smaller than the main asteroid belt, librates around Jupiter's stable L4 and L5 Lagrange points, 60 deg ahead and behind Jupiter. It is thought that these bodies were captured into these orbits early in solar system history, but any capture mechanism must also explain why the Trojans have an excited inclination distribution, with some objects reaching inclinations of 35°. The Trojans themselves, individually and in aggregate, also have spectral and physical properties that appear consistent with many small bodies found in the outer solar system (e.g., irregular satellites, Kuiper belt objects). In this review, we assemble what is known about the Trojans and discuss various models for their origin and collisional evolution. It can be argued that the Trojans are unlikely to be captured planetesimals from the giant planet zone, but instead were once denizens of the primordial Kuiper belt, trapped by the events taking place during a giant planet instability. The Lucy mission to the Trojans is therefore well positioned to not only answer questions about these objects, but also about their place in planet formation and solar system evolution studies.

# 1 Introduction

Lucy's mission to Jupiter's Trojan asteroids (hereafter Trojans) will examine one of the last unexplored small body reservoirs in the Solar System (Levison et al. 2021). By visiting at least eight Trojan bodies over a 12-year traverse, Lucy will probe the origin and evolution of these singular objects, which in turn may tell us about planetesimal and planet formation processes in the outer solar system, the effects of giant planet migration on small body populations, and how such worlds have experienced collisional evolution. The Trojans also potentially set constraints on a dramatic phase of solar system history when the giant planets may have experienced a dynamical instability after the solar nebula had dissipated. By understanding the Trojans, we can test numerical models that suggest the giant planets formed on different orbits than they have today and that Neptune's migration through the primordial Kuiper belt injected bodies into numerous stable zones, ranging from the main asteroid belt to the irregular satellites of the giant planets to specific regions of the transneptunian populations.

At this time, researchers have limited information about the composition and physical characteristics of the Trojans and know nothing about their cratering history (see chapters by Mottola et al. and Marchi et al., this volume). Ideally, an investigation of these properties will provide vital clues concerning their provenance location (e.g., near Jupiter's orbit, or within the primordial Kuiper belt) and what the nature of the solar nebula was like at those locations when they formed. In order to interpret the data provided by Lucy's observations, we need be able to place the Trojans into a science context. Here we briefly summarize what is known of these worlds. The interested reader should also seek out two excellent comprehensive reviews of the Trojans by Marzari et al. (2002) and Emery et al. (2015) as well as the other Trojan chapters in this volume.

# 2 Trojan Characteristics

## 2.1 Orbits

Jupiter Trojans (hereafter Trojans) are populations that orbit around the L4 and L5 Lagrange points of Jupiter, which are located 60° in front of and behind Jupiter, respectively. They are contained within Jupiter's 1:1 mean motion resonance, with Jupiter having a semimajor axis of 5.2 au (**Fig. 1**). Details on their dynamical status can be found Marzari et al. (2002) and Emery et al. (2015). The regions themselves are relatively small, and it is unlikely that the currently observed bodies formed *in situ*. Instead, most experts argue that the Trojans were captured there during the earliest times of the Solar System, possibly when Jupiter was still forming and/or migrating. Various formation and capture scenarios will be discussed in Sec. 3.

PLACE FIGURE 1 HERE

Most Trojans reside on stable orbits that librate around L4 and L5, but a small fraction is unstable over ages shorter than the lifetime of the Solar System (Levison et al. 1997; Tsiganis et al. 2005; Di Sisto et al. 2014; 2019; Holt et al. 2020). Some unstable bodies are also sizeable; examples include (1173) Anchises, which is 100 km in diameter, and (1868) Thersites, which is 70 km in diameter (Tsiganis et al. 2000; Horner et al. 2012). These Trojans were likely captured on orbits near the stable zone boundary, and that means they will slowly diffuse out of the Trojan zone on billion-year timescales (e.g., Robutel and Gabern 2006). The remaining unstable Trojans are likely to be temporary captures from the Jupiter family comet population. The regions around L4 and L5 are dynamically "sticky", and objects that evolve to similar orbital conditions can find themselves caught for an extended time. It is also possible that some unstable Trojans reached their orbits by one of the following mechanisms: (i) the same mechanism that captured the Trojans (see Sec. 3), (ii) collisions, and/or (iii) some long-term dynamical process such as Yarkosky thermal drift or outgassing. Regardless, it can be shown that most Trojans many tens of kilometers in diameter or larger are on stable orbits, so they we can presume they are survivors of the same process that created the Trojans in the first place. We can use these bodies to evaluate Trojan origin scenarios.

Arguably the most remarkable dynamical property of the Trojans are their inclinations, which cover a fairly uniform range from 0° to 35° in both the L4 and L5 region, with a median inclination of 10° (**Fig. 1**). The eccentricities of the Trojans tend to be $e < 0.15$, values required to remain within the stable L4 and L5 zones (e.g., Levison et al. 1997; Nesvorný and Dones 2002). The dynamical excitation of the Trojans presents a substantial challenge to all origin models. For example, not only must a proposed capture mechanism place comparable numbers of bodies near Jupiter's L4 and L5 zones, but it must also explain how the populations obtained high inclinations (i.e., the populations were either excited within L4/L5 in similar ways after capture or they were implanted with an excited inclination distribution). Over the last several decades, many origin scenarios have failed the test of explaining Trojan inclinations.

## 2.2 Population

The largest Trojan is (624) Hektor, whose diameter is $D{\sim}250$ km (Marchis et al. 2014), while smallest observed Trojans to date are likely to be a kilometer or so across (e.g., Yoshida and Terai 2017). Overall, the combined Trojan populations of L4 and L5 are considerably smaller than the population in the main asteroid belt, though how much smaller depends on how one evaluates the size frequency distributions (SFD) of each population, which do not have the same shape.

For example, both the main belt and the Trojans have a "bump" in their cumulative SFDs for $D \sim 100$ km objects (i.e., the power law slope of the SFD is substantially steeper for $D > 100$ km bodies and is substantially shallower for $D < 100$ km objects). This bump means that most of the mass contained within the connected slopes of the SFD are in those size bodies. Indeed, it can be argued that on that basis, $D \sim 100$ km objects are a likely to be a characteristic size produced by planetesimal formation processes (Bottke et al. 2005; Morbidelli et al. 2009). Using this criterion, it can be shown that the Trojans are approximately ten times less massive than the main belt population; ~20 or so bodies in the Trojans vs. ~200 in the main belt (Bottke et al. 2020; Marschall et al. 2022). Other estimates suggest this ratio holds fairly well down to km-sized bodies as well; the Trojans are projected to have $\sim 10^5$ bodies with $D > 1$ km, while the asteroid belt has $\sim 10^6$ such objects (e.g., Nakamura and Yoshida 2008; Szabó et al. 2007; Bottke et al. 2020).

The relative size of the L4 and L5 Trojan populations is debated. Dynamically, there is no particular difference between the two regions that would lead to a pronounced asymmetric population by Trojan escape (e.g., Holt et al. 2020). While it is possible that a putative Trojan capture mechanism may have injected more bodies into one cloud than the other, it is not clear the population data supports this idea. The population of $D > 100$ km bodies, where most of the mass of the Trojans resides, contains 9 objects in both L4 and L5. Agreement between the populations would probably be acceptable for values of $9 \pm 3$. Accordingly, for the bodies least likely to be affected by collisional evolution (e.g., Bottke et al. 2005), it would appear that the two populations are comparable to one another.

If we examine $D > 10$ km bodies, the ratio of L4 to L5 objects moves to $1.4 \pm 0.2$ (Grav et al. 2011), values within the fractional errors given above for the $D > 100$ km bodies. These results are consistent with both populations being statistically comparable, but it is also possible that some modest asymmetry may exist (e.g., Marschall et al. 2022). This putative discrepancy will be discussed in Sec. 3 in the context of Trojan origins.

## 2.3 Size Distribution

The current Trojan SFD gives us crucial information about the initial Trojan population. Both the L4 and L5 swarms have similar SFDs and follow a broken power law for observed objects (**Fig. 2**). The largest Trojans follow cumulative power law slope $q = -5$ for $D > 100$ km bodies (e.g., Jewitt et al. 2000; Fraser et al. 2014) and a cumulative slope of approximately $q \sim -2.1$ between 10 km and 100 km (right panel of **Fig. 2**). The shape of the Trojan SFD for objects larger than several tens of kilometers is similar to that found among Kuiper belt objects, and it has been suggested that both have originated from the same source population (Morbidelli et al. 2009a; Fraser et al. 2014; Emery et al. 2015).

PLACE FIGURE 2 HERE

For smaller bodies, Grav et al. (2011) found a cumulative slope of $q = -2.1$ between ~10 and 100 km for both the L4 and L5 swarms, values in agreement with Jewitt et al. (2000). Pencil beam studies by several groups (Yoshida and Nakamura 2008; Wong and Brown 2016; Yoshida and Terai 2017) indicate that the same approximate slope continues to sizes smaller than 10 km. Beyond those sizes, however, the nature of the SFD becomes less clear.

Several observational studies suggest an inflection point exists in the SFD somewhere between a few kilometers and 10 kilometers, with objects smaller than the inflection point following a SFD with a shallower power law slope. For example, Yoshida and Nakamura (2008) reported a break in the SFD for absolute magnitude $H \sim 16$ in the L4 cloud. Follow up work by Wong and Brown (2016) confirmed the break but found it at $H \sim 15$. For an albedo of 0.07 this would correspond to a change in slope between 3 km and 10 km. Using HyperSuprime Cam on the Suburu telescope, Yoshida and Terai (2017) and Uehata et al. (2022) found comparable results to these studies, but also argued that a single power law of $q = -1.85 \pm 0.05$ could be fit to L4 and L5 Trojans between $13.6 < H < 17.4$, approximately 2-10 km diameter bodies, assuming an albedo of 0.07. Future observations by the Dark Energy telescope, Vera Rubin telescope, or the NEO Surveyor mission may be needed to settle the issue of what happens to the SFD of smaller Trojans.

## 2.4 Spectral Properties

Observations of the Trojans show that they are dominated by objects that have the broad appearance of dormant comets. Visible spectroscopy indicates Trojans have featureless spectra in visible and near-infrared wavelengths, with spectral slopes ranging from neutral to moderately red (e.g., Dotto et al. 2006; Fornasier et al. 2007; Melita et al. 2008). Using taxonomy based on main belt asteroids, Trojans mainly fall within the D- and P-type categories (e.g., Emery et al. 2015). In this nomenclature, D-types are considered "red" and the P-types are considered "less red" (see also Brown et al. 2014 and Wong et al. 2014). The differences in colors may be related to where these objects formed within the solar nebula, how they evolved prior to capture within the Trojans, with some passing closer to the Sun than others, collisional evolution within the Trojan population over the last several billion years, or possibly all of these issues combined. Our discussion of Trojan colors relative to other populations (e.g., Kuiper belt) can be found in Sec. 5.3.

A small fraction of the Trojans are C-types. Most of these bodies are in the Eurybates asteroid family located within the L4 population (**Fig. 1**). This family will be discussed more below in Sec. 5.2. It is possible that their taxonomic type is related to their collisional disruption, with the Trojans looking different on the inside than on the outside (Wong and Brown 2016; 2017a).

Trojans also tend to have low albedos that are consistent with D- and P-type asteroids in the main asteroid belt or dormant comets (Grav et al. 2011) as well as the majority of C-complex asteroids (Masiero et al. 2013). Barucci et al. (2002) found albedos for the Trojans they observed between 0.03 and 0.07, while NEOWISE observations found albedos of $0.07 \pm 0.03$ across all sizes (Grav et al. 2011; see also Emery et al. 2015).

**2.5 Rotation and Physical Properties**

Until recently, information on the rotation rate distribution of Trojans was limited to a few tens of Trojans. The interested reader should see the review and references in Emery et al. (2015) to find out more about the individual rotational light curve studies used to gather these data. In the section, we briefly discuss recent work determined from space-based telescopic surveys to set the context for later discussions in this chapter on Trojan formation and collisional evolution. The topics mentioned here are discussed in more detail by Mottola et al. (this volume).

The spin periods for 56 Jupiter Trojans, many of them $15 < D < 50$ km, were determined using Kepler telescope photometry by Szabó et al. (2017) and Ryan et al. (2017). These data were gathered during their K2 Campaign 6 that lasted from July 14, 2015 to September 30, 2015. A crucial reason to use Kepler data is that many objects were observed for months, and as such exclude observation selection effects against slow rotators. They found that eight out of 53 Trojans, or ∼15%, have extremely long spin periods ($P \geq 100$ hours). A follow-up study by Kalup et al. (2021) extended their analysis of Kepler data to 101 Trojans. They showed there is a dichotomy in Trojan spin periods, with the division taking place near 100 hours; 25% had $P \geq 30$ hours, and 12% were very slow rotators with $P \geq 100$–600 hr. Nesvorný et al. (2020a) argued that the easiest explanation for this excess was that prior to implantation within L4 or L5, these bodies were nearly equal-sized binaries that had tidally evolved toward a synchronous state. The process of capture likely led to their companions being stripped away, leaving behind a sizeable set of very slow rotators.

The Nesvorný et al. (2020a) proposal, if true, has several implications. First, equal-sized binaries may have been a common outcome of planetesimal formation wherever Trojans actually formed. Second, collisional evolution among the Trojans was not enough to substantially alter the spin state of slow rotators stripped of their companion. This scenario may explain why (11351) Leucus, a $D \sim 35$ km target of the Lucy mission (see Levison et al., this volume), has a rotation period of 445.73 hours (French et al. 2015; Buie et al. 2018; Kalup et al. 2021). Third, there may be many binaries among the Trojans today. Both Szabó et al. (2017) and Kalup et al. (2021) use the slow rotation periods and high lightcurve amplitudes of their observed Trojans to argue for a ∼25% rate of binaries within their samples.

As a second example, the publicly-released photometry data of the Zwicky Transient Facility was used by Schemel and Brown (2021) to obtain color, phase parameter, absolute magnitude, and amplitude of rotation measurements for 1049 Trojan asteroids. Overall, they found that the distribution of rotational light curve amplitudes is relatively constant with asteroid size, and they argued it was not distinguishable from that of main belt asteroids of similar sizes.

This result can be compared with the interpretation of observations taken from the Asteroid Terrestrial-impact Last Alert System (ATLAS). Using sparse photometric data for 863 L4 Trojans and 380

L5 Trojans, McNeill et al. (2021) argued that L4 asteroids are more elongated than the L5 asteroids. The group claimed this could be a byproduct of greater collisional evolution within the L4 swarm, which may have slightly more $D > 10$ km bodies (Grav et al. 2011). We suspect this scenario is unlikely, mainly because the collision probabilities within L4 or L5 are comparable to one another (e.g., Nesvorný et al. 2018). The discrepancy between McNeill et al. (2021) and Schemel and Brown (2021) has yet to be resolved.

## 3  Origin Scenarios for Trojans

### 3.1 Review of Trojans origin models

One reason that the Trojans hold our fascination is because their origin story has defied conventional explanations for decades. There is no dynamical pathway for a heliocentric body to reach a stable orbit near L4 or L5 within the context of the three-body problem (i.e., Sun, Jupiter, and an asteroid). Either the Trojans had to form *in situ*, which is unlikely, or some kind of extra perturbation was needed to capture them. This has led to a plethora of origin scenarios for the Trojans. We refer the reader to the comprehensive review of Trojan origin scenarios provided by Marzari et al. (2002) and Emery et al. (2015).

For example, a natural idea is that the Trojans are remnant planetesimals from the Jupiter zone that were captured in L4 or L5 via gas drag. As Jupiter was forming, it was proposed that gas drag from the solar nebula could have provided the kind of dissipative force needed to overcome the limitations of the three-body problem (e.g., Yoder 1979; Peale 1993). Alternatively, mutual collisions between planetesimals might have also managed to trap objects that happened to be lingering near L4 or L5 (Shoemaker et al. 1989).

Additional possibilities come from the idea that Jupiter's growth or migration would lead to a shift in the location and properties of L4 and L5, leading to capture among planetesimals that happened to be at the right place at the right time. An example of such a model would be an increase in the growth of Jupiter via accretion of planetesimals, protoplanets, and/or gas (Marzari and Scholl 1998a,b; Fleming and Hamilton 2000). Jupiter's migration in a gas disk could also lead to Trojan capture, provided the migration was smooth enough that planetesimals trapped along the way would remain trapped (e.g., Lykawka et al. 2009; Pirani et al. 2019a). It should be noted that the Pirani et al. (2019a) mechanism shows that an inward migrating Jupiter will capture more Trojans within L4 than L5. If it can be shown that the population asymmetry favoring L4 over L5 is real and statistically robust, this result would represent evidence in favor of their origin model (though the Pirani et al. (2019a) mechanism has other issues, as will discussed below).

Our expectation for these models is that the Trojans were planetesimals that were formed as a byproduct of gas and dust concentration mechanisms in the solar nebula. Given this condition, it is likely that many planetesimals would be residing near the ecliptic if they happened to be captured while the solar nebula was still in existence. This leads to an Achilles' heel for many capture models, however, in that most new Trojans would retain low inclinations once the solar nebula dissipated.

So far, no mechanism has yet been identified that can substantially increase the inclinations of captured Trojans to those matching observations, which weakens the scenario suggested by Pirani et al. (2019a). Secular resonances produced by other worlds are unable to do it, nor are protoplanets that flyby and gravitationally interact with bodies already trapped within L4/L5 (e.g., Marzari et al. 2002; Pirani et al. 2019b). It is conceivable that planetary embryos were temporarily captured within L4 and L5, thereby exciting the Trojans, but numerical simulations indicate such large bodies are unlikely to both (i) produce the appropriate inclination distributions in each cloud and (ii) escape prior to the present day (Pirani et al.

2019b). Instead, such bodies would most likely still be there and would thus be easily observable. Accordingly, the inclination distribution of the Trojans makes it difficult to accept any of the proposed capture models discussed above.

The Trojan origin models best able to explain the inclination constraint are those that assume that the planetesimals were excited prior to capture, and that the capture mechanism itself involves the migration of multiple giant planets along with Jupiter.

## 3.2 Review of Trojan origin models involving a giant planet instability

As discussed above, a good Trojan origin model requires several components:

T1) A planetesimal population that has been dynamically excited prior to capture,
T2) A mechanism that involves Jupiter's migration, since it has the ability to modify the shape of the L4 and L5 zones, and
T3) A scenario capable of placing an appropriate number of objects into L4 and L5 with the appropriate SFDs and physical/spectral properties while also satisfying other solar system constraints.

For T1, while it is possible that planetesimals in the giant planet zone became excited during the formation of the giant planets themselves, and thus could have been captured while gas was still around, it seems likely that such a scenario would predominantly produce low inclination Trojans. This is not to say that zero Trojans were captured at this time, but something had to happen to them, otherwise this putative population would skew the current Trojans toward low inclinations. In addition, based on main belt modeling predictions and constraints, our expectation is that captured planetesimals from the giant planet zone would look much more like C-type asteroids than D- and P-type asteroids (e.g., Walsh et al. 2011).

Taken together, it seems probable that the Trojans were dynamically excited after the solar nebula had dissipated than before it was gone. This means that to satisfy T2, we need Jupiter's migration to also occur after the solar nebula had been eliminated. Finally, for T3, we need to destabilize an excited small body population that can satisfy the physical constraints for the Trojans.

### 3.2.1 The giant planet instability

At present, the best model that can explain such behavior is characterized as the so-called "Nice model", an umbrella term for a suite of models that involves dramatic giant planet migration occurring after the elimination of the gas disk (Tsiganis et al. 2005; see Nesvorný 2018 for a review). Here we describe their basics.

After the gas disk disperses, our system of giant planets was in a different configuration than the one seen today. Gas accretion left them on nearly circular coplanar orbits between ~5 and ~20 au, with most/all locked in mutual mean motion resonances with one another. Beyond that zone, a massive primordial Kuiper belt existed just beyond the original orbit of Neptune, with an estimated comet population of ~20 Earth masses or more that stretched well beyond 20 au. This configuration, however, eventually went unstable. The most natural explanation for the dynamical structure of the present-day Kuiper Belt is that the outermost giant planets radially migrated through the primordial Kuiper belt in a violent exchange of orbital energy and angular momentum. This led to a reorganization of the giant planets in what is now called a "giant planet instability" (e.g., Tsiganis et al. 2005; Nesvorný and Morbidelli 2012; Nesvorný 2018).

As Neptune entered into the primordial Kuiper belt, it excited the objects residing there via resonant interactions. The subsequent migration of Neptune through this population caused most of the bodies to be flung throughout the Solar System. It led the primordial Kuiper belt to lose a factor of ~1000 in population, with roughly this number of Pluto-sized bodies -- and enumerable smaller bodies -- scattered into the giant planet zone and/or out into a scattered disk of comets associated with Neptune. The interaction between Neptune and Pluto-sized bodies made the migration grainy rather than smooth, and this in turn affected the capture of Kuiper belt objects into Neptune resonances, explaining observations (Nesvorný and Vokrouhlický 2016; Nesvorný et al. 2016; Kaib and Sheppard 2016; Lawler et al. 2019).

The number of giant planets gravitationally interacting with one another at the time of the instability is unknown. Dynamical studies suggest systems with 5 or 6 giant planets (e.g., 3 or 4 Uranus/Neptune-size bodies) have greater success reproducing dynamical constraints across the solar system than those that start with 4 giant planets (Nesvorný 2011; Nesvorný and Morbidelli 2012; Batygin et al. 2012). In the most successful cases, a Neptune-sized body interacting with Jupiter causes it to migrate slightly inward via numerous tiny jumps. These events may help describe the dynamical structure of the asteroid belt and terrestrial planet system, and is referred to as the "Jumping Jupiter" scenario (e.g., see review in Morbidelli et al. 2015).

### 3.2.2 Trojan capture during the giant planet instability

From this context, we can examine several different but fairly related capture scenarios for the Trojans. For example, in the original Nice model, Morbidelli et al. (2005) proposed that gravitational interactions with refugees from the primordial Kuiper belt led Jupiter and Saturn to enter into their mutual 1:2 mean motion resonance. This resonance crossing had several effects. First, it triggered the giant planet instability and destabilized the outer solar system, with Uranus and Neptune eliminating most of the primordial Kuiper belt. This caused large numbers of objects, with a range of inclinations (see below), to be sent to the vicinity of Jupiter. Second, just after Jupiter and Saturn crossed their mutual 1:2 resonance, Jupiter's L4 and L5 regions became unstable for a short time, allowing refugee Kuiper belt objects to both enter and exit. Further giant planet migration closed the doors on the L4 and L5 regions, leaving the objects left inside those regions on stable orbits.

An issue here is that in this model, and in all models using a giant planet instability, Trojans are captured Kuiper belt objects. The model prediction is that the size distribution, colors, and physical nature of the Trojans should match what we know of the Kuiper belt and other populations that were captured at the same time by the same process, unless they are affected by subsequent evolutionary processes (e.g., collisions). This implies that Trojans are siblings of D- and P-type captured in the central and outer main belt, Hilda asteroids, irregular satellites, and Neptune Trojans (Vokrouhlický et al. 2016; see Nesvorný 2018 for a review).

This initial version of the Nice model, and its associated Trojan capture mechanism, has increasingly become disfavored. Numerical simulations and studies of exoplanets have shown that the giant planets likely evolved into mutual mean motion resonances prior to the end of the solar nebula (Morbidelli et al., 2007; Walsh et al., 2011; see Morbidelli 2013 for a review). This would suggest that the giant planet instability was not triggered when Jupiter and Saturn entered the 1:2 mean motion resonance, but rather when some of the planets broke out of resonant lock. Numerical studies show this is a common occurrence taking place shortly after the solar nebula has dissipated. In effect, the initial giant planet configuration is stable with gas, but relatively unstable without it, particularly when interacting with a large primordial Kuiper belt that may be losing bodies to the giant planet zone.

In this alternative version of the giant planet instability, giant planet encounters with one another cause them to change orbits, moving their L4 and L5 stability zones in a sudden fashion. This means that a small fraction of wandering Kuiper belt objects will have a "dynamical net" thrown over them, capturing them in stable orbits if they happen to be on the right orbits. The advantage of this model is that it better reproduces the observed distribution of Trojans, and it allows the stochastic nature of Trojan capture events to potentially produce a modest asymmetry in the populations within L4 and L5 (Emery et al. 2015). This model also favors Jupiter interacting with an extra Neptune-sized body that is eventually thrown out of the solar system (Nesvorný 2011; Nesvorný and Morbidelli 2012; Batygin et al. 2012; Nesvorný et al. 2013; Vokrouhlický et al. 2016).

The probability that primordial Kuiper belt objects will be captured in the Trojan population at this time is approximately one in a million (i.e., $\sim 5 \times 10^{-7}$) (Nesvorný et al. 2013). This value is comparable to the capture probability of similar objects in the Hilda asteroids (i.e., those in the 3:2 mean motion resonance with Jupiter) and the irregular satellites of the giant planets, but it is several times smaller than the probability of being captured in the main belt (e.g., Vokrouhlický et al. 2016). Accordingly, we would expect more D- and P-types in the main belt than the Trojans, and the largest body in the former should exceed that of the latter, both which are true (Vokrouhlický et al. 2016). Overall, these models reproduce the numbers and sizes of the largest P- and D-type asteroids in each of the aforementioned populations except the irregular satellites, which are likely experienced exceptional levels of collisional evolution (Bottke et al. 2010; Vokrouhlický et al. 2016).

This model also predicts that the Trojans, the dynamically hot Kuiper belt population, and the scattered disk are all from the same primordial Kuiper belt (Levison et al. 2008). From existing observations, it appears the SFD of the dynamically hot Kuiper belt is consistent with this idea; both show a Trojan-like like bump near 100 km and a similar shape for the observed bodies (Morbidelli et al. 2009; Adams et al. 2014; Fraser et al. 2014). They may also be a match with the largest bodies in the scattered disk SFD, which is the source of the ecliptic comets, but observational incompleteness among scattered disk bodies and ecliptic comets makes this prediction difficult to test at this time.

### 3.2.3 Details on the evolution of the primordial Kuiper Belt

The results presented above imply that the inclination distribution of the Trojans was obtained from objects that came from a dynamically excited primordial Kuiper belt that was further excited prior to capture by giant planet interactions. Accordingly, in order to interpret the nature of present-day Trojans, we need to understand the initial conditions of the primordial Kuiper belt, how and when this population became excited, and what happened to ejected objects en route to being captured in the Trojan population. On this basis, the properties of the present-day Kuiper belt, scattered disk and even ecliptic comets provide us with insights into the nature of the Trojan population, and vice versa.

Since Tsiganis et al. (2005) and Morbidelli et al. (2005), there have been many advances in our thinking about the nature and timing of the giant planet instability and the capture of the Trojans. More recent numerical simulations indicate the timing of the giant planet instability is controlled by the unknown size of the gap between Neptune's initial orbit and the primordial Kuiper belt's inner edge (Nesvorný et al. 2018). The smaller the gap, the earlier the instability. If the gap is < 1.5 au, there is effectively no delay in the onset of Neptune's migration, and it would take place shortly after dissipation of the solar nebula. That might explain why the Trojans appear to have experienced limited collisional evolution, as we will discuss below.

The initial state of the primordial Kuiper belt is thought to have been dynamically cold, a necessary condition for planetesimal formation. The objects can be assumed to have had low eccentricities ($e < 0.1$) and inclinations ($i < 10°$) prior to Neptune's outward migration. The disk itself stretched from near 24 au, a distance modestly outside Neptune, to at least 30 au, the current semimajor axis of Neptune, and probably beyond. In the original Nice model, it was assumed that Neptune would migrate through the disk until it reached the outer edge and/or the spatial density of the objects dropped low enough to terminate planetesimal-driven migration (Tsiganis et al. 2005). If the former, the edge would need to be located near 30 au, but this scenario has difficulty explaining the cold classical Kuiper belt, whose outer edge is at ~45-50 au.

Concentrating on the latter scenario, models can now show that the cold classicals can be explained by assuming the spatial density of objects in the primordial Kuiper belt gradually decreased away from the Sun (Nesvorný et al. 2020). This idea is plausible, with photoevaporation potentially reducing the gas density from the outside-in and thereby limiting planetesimal formation in the cold classical region and beyond (Carrera et al. 2017). That would allow Neptune to migrate through the population until there was insufficient mass for it to move any further. The orbital distance where Neptune ground to a halt would be 30 au, its current semimajor axis. From here, the only objects left outside of Neptune's orbit would be those captured in mean motion resonances with Neptune by Neptune's outward migration, destabilized objects that would eventually be ejected from the primordial Kuiper belt by Neptune perturbations (i.e., the scattered disk), destabilized objects scattered by a more eccentric Neptune that were left stranded when Neptune evolved to its current eccentricity (i.e., the detached scattered disk), and those so far away from Neptune that they were not dynamically affected by Neptune's migration.

The last population is a good fit with the properties of the cold classical Kuiper belt. It is home to objects like Arrokoth, a body that was likely formed *in situ* (Batygin et al. 2011; Parker and Kavelaars 2012; Fraser et al. 2014; McKinnon et al. 2020). Note that recent models suggest Arrokoth and other KBOs were produced when clouds of small particles experienced gravitational collapse (see Sec. 5.1). The unaffected nature of the cold classical Kuiper belt would also explain why it has a high fraction of well-separated binaries relative to the fraction in the dynamically hot Kuiper belt.

As Neptune migrated outward through the disk, it traveled slowly enough to dynamically excite the eccentricities and inclinations of objects located modestly ahead of it (Nesvorný and Vokrouhlický 2016). Most were destabilized and sent to the giant planet zone. This evolution probably explains the inclinations of the Trojan swarms. Some of the destabilized bodies from the primordial Kuiper belt reaching the giant planet zone managed to be in the right place when Jupiter was having encounters with another giant planet, and this led to their capture into many stable zones, such as the central and outer asteroid belt, the 3:2 and 1:1 mean motion resonances with Jupiter, and in distant orbits around Jupiter itself. These worlds now dominate the D- and P-type populations seen in the asteroid belt, Hildas, Trojans, and irregular satellites of Jupiter.

Finally, some objects destabilized by Neptune were sent to a scattered disk associated with Neptune's orbit; this region is now the primary source of our Centaurs and ecliptic comets (e.g., Duncan and Levison 1997; Nesvorný et al. 2019b). Other bodies were passed down the other giant planets, where they were ejected. A modest fraction of this population managed to reach Oort cloud, potentially explaining the origin of these bodies (e.g., Vokrouhlický et al. 2019), though some Oort cloud comets may also have been captured from passing stars in our primordial star cluster (Levison et al. 2010). These associations mean that the Trojans have many siblings across the solar system. Some of these bodies are still in cold storage, while others are just coming out of it as they approach the Sun. Many Trojan siblings are also collisionally evolved from their time in the excited primordial Kuiper belt.

## 4 Collisional and Physical Evolution of Trojans Prior to Capture

In this section, we consider the physical and collisional evolution of the Trojans. By adopting the above origin scenario, we adopt the idea that most or perhaps all of the Trojans were once members of the primordial Kuiper belt. That means many Trojans have physical histories that predate their capture in Jupiter's L4 and L5 Lagrange points. We divide this history into several stages.

### 4.1 Planetesimal and binary formation in the primordial Kuiper belt

In stage one, we consider the formation of Trojans as icy planetesimals in a primordial Kuiper belt that was still dynamically cold. Theoretical work and numerical simulations indicate that planetesimals were likely formed by concentrating small grains in protoplanetary disks. Specifically, the streaming instability model suggests that pebbles, defined as mm- to cm-sized particles, move toward high pressure zones in the solar nebula, and are aerodynamically collected into self-gravitating clouds (Youdin and Goodman 2005; see review by Johansen and Lambrechts 2017). When a sufficient spatial density of pebbles is reached, the particle clouds undergo gravitational collapse and form into short-lived disk structures from which planetesimals accrete. The size of the bodies formed can have a range of sizes, but the most common diameter of the largest planetesimals created is ~100 km (Klahr et al. 2020; 2021). They may even produce numerous equal size components that coalesce together, as has been suggested to explain the mounds on Arrokoth (Stern et al. 2023).

Numerical simulations indicate that particle/gas clouds with low angular momentum will frequently produce a single largest planetesimal with a size near ~100 km, but those with higher degrees of angular momentum will often lead to the formation of equal-sized binary components orbiting one another at substantial distances (Nesvorný et al. 2019; 2021). These latter binaries are consistent with observations of many Kuiper belt binaries; the components are near 100 km in diameter, they are nearly equally sized components, they can have wide separations, and they have similar colors. These types of binaries are most common in the cold classical Kuiper belt (i.e., 40 out of 65 investigated are such binaries) (Noll et al. 2020). They are less common elsewhere, but that could be a consequence of the bodies undergoing collisional and dynamical evolution in their source populations during the giant planet instability and Neptune's migration through the primordial Kuiper belt.

These same planetesimal formation events frequently produce and eject a number of smaller bodies as well (Nesvorný et al. 2019; 2021). In some cases, the smaller objects form from flattened disks spun off from larger clusters of particles, leading to shapes that may resemble hamburgers. These kinds of origin conditions may provide a readily explanation for the flattened appearance of the two joined lobes of Arrokoth (e.g., McKinnon et al. 2020) and that of the Lucy target Polymele (see chapter by Buie et al., this volume).

For example, in the hot classical Kuiper belt today, many binaries with a large primary also have a satellite less than half the size of the primary (though binaries with smaller equal-size components also exist) (Nesvorný and Vokrouhlický 2019). It is possible these satellites formed from impact-generated disks after their primary was struck during stage one collisions (Canup 2005; Leinhardt et al. 2010; Sekine et al. 2017). Some evidence supporting this comes from Sekine et al. (2017), who argue that high-velocity giant impacts between KBOs my produce global/hemispherical darkening and reddening, thereby explaining the color variations observed among large KBOs. If true, reasons the aforementioned binaries are not found in the cold classical population may be because that population was smaller, and therefore experienced fewer stage one collisions, or because they are simply harder to detect.

Our expectation is that the fraction of well-separated binaries with equal-sized components in the primordial Kuiper belt once rivaled the fraction seen in the cold classicals. Moreover, the destabilized Kuiper belt population and the hot classical Kuiper belt probably came from the same region of the primordial Kuiper belt, so we can deduce that some of these bodies were in binaries formed from impact-generated disks. The idea that many Trojans were previously in some kind of binary is consistent with observations from the Kepler spacecraft that ∼15% of Jupiter Trojans have very slow rotation (spin periods > 100 hr; see Sec. 2.5). Models indicate these slow rotators could have had their rotational angular momentum taken away by gravitational interactions with a satellite, with the satellite then stripped away via gravitational interactions with a giant planet prior to Jupiter capture or destroyed by collisions (Nesvorný et al. 2020).

These ideas fit well with the diversity of the Lucy targets, of which three are known to have satellites. For example, Leucus, which is $63.8 \times 36.6 \times 29.6$ km D-type Trojan, is a very slow rotator with a spin period of 445.68 hours (see chapter by Mottala et al., this volume). It is likely it was one component of a stripped binary. Patroclus, which we will discuss in Sec. 4.2, and Polymele both have satellites that were probably created and/or captured during the planetesimal formation process. We consider it less likely that collisions in the primordial Kuiper belt created Queta, the small satellite of the target Eurybates (see Sec. 5.3).

These planetesimal formation mechanisms, along with accretion between planetesimals and pebble accretion, led to the formation of KBOs ranging from sub-100 km planetesimals to Pluto-sized bodies (and perhaps beyond). The net mass of the primordial Kuiper belt prior to collisional evolution was probably ~20-30 Earth masses, and it contained the order of ~$10^3$ Pluto-sized bodies. Once created, the Pluto-sized bodies would have modestly excited the surrounding planetesimals, possibly enough to produce some degree of collisional grinding at slow impact velocities (i.e., tens to hundreds of m/s). The magnitude of stage one collisional grinding, however, is unknown, and our knowledge of how these collisions influenced the primordial Kuiper belt SFD and Kuiper belt binaries is limited.

It is also conceivable that stage one collisional evolution may be responsible for the observed cumulative power law slope of $q = -2.1$ found among observed Trojans with $D < 100$ km. The idea is that low velocity collisions in the primordial Kuiper belt may produce sufficient grinding to create a Dohnanyi-type size distribution (Dohnanyi 1969; O'Brien and Greenberg 2013; see Supplemental Figure 4 in Nesvorný et al. 2018). The shape of this SFD would then be transferred to small body reservoirs across the solar system, including the Trojans. This scenario may be testable by examining the physical nature of Trojans (e.g., what evidence to Trojans show for very early collisional evolution history in addition to their observed craters).

**4.2 Evidence for an early instability from the Patroclus binary**

Using the ideas presented above, we now consider an application of stage one collisional evolution. The most prominent example of an equal-sized binary in the Trojan population is (617) Patroclus and its satellite Menoetius, which we hereafter call the P-M binary. Both P-M components are larger than 100 km, and they are separated by 670 km. Located in Jupiter's L5 cloud, they will be visited by the Lucy spacecraft in 2033. As suggested above, the P-M binary likely formed in the primordial Kuiper belt as a byproduct of planetesimal formation, and its existence in the Trojans provide insights into the nature of the primordial Kuiper belt, a possibility recognized by Nesvorný et al. (2018). The P-M binary had to endure stage one collisional grinding in the primordial Kuiper belt between 20-30 au, collisional evolution as a destabilized binary en route to the Trojans, and disruption events in the Trojan population over the last ~4.5 Gyr. Moreover, it had to avoid being stripped by encounters with a giant planet prior to capture in L5.

By putting together a collisional and dynamical model for the primordial Kuiper belt, and assuming a test population made of P-M-like binaries, Nesvorný et al. (2018) estimated the attrition rate of these bodies over time. Their results showed that if the giant planet instability occurred immediately after the dissipation of the solar nebula, and there were no stage one collisions, roughly 60% of their model P-M binaries would survive (i) stage two and stage three collisions (i.e., collisions from an excited primordial Kuiper belt and collisions from the Trojans over 4.5 Gyr, respectively) and (ii) encounters with giant planets capable of producing binary stripping. When the giant planet instability occurred at later times, allowing stage one collisions to become more important, the survival rate of the P-M binaries dropped precipitously. In other words, the earlier the giant planet instability, and the faster stage one collisions come to an end, the higher the odds that a P-M binary would survive to become a Trojan.

Ultimately, Nesvorný et al. (2018) predicted that the giant planet instability had to start within ~100 Myr of the first solids, but they found the most favorable results when migration started shortly after the dissipation of the solar nebula. This argues against late migration for Neptune, which is problematic for other reasons (e.g., it is difficult to keep giant planets in a stable resonant configuration for long time periods without a gas disk, as discussed by Nesvorný (2018); late giant planet migration can produce dynamical damage to fully formed terrestrial planets; Agnor and Lin (2012); Kaib and Chambers 2016; Clement et al. 2018; Nesvorný et al. 2021). It also means there are hard limits to how much stage one collisional evolution could have occurred.

### 4.3 Insights into small Trojans and KBOs from craters on Charon and Arrokoth

At some unknown but presumably early time, Neptune entered the primordial Kuiper belt, migrated through it, and started stage two collisional evolution. Here the primordial Kuiper belt became dynamically excited, and for the first time, the objects observed by the New Horizons spacecraft, namely Pluto-Charon and Arrokoth, were capable of being hit by a large fraction of the excited primordial Kuiper belt population. Collision velocities on Pluto-Charon and Arrokoth were raised to 2.2 and 0.7 km s$^{-1}$, respectively, and numerous cratering events should have taken place as the destabilized portion of the primordial Kuiper belt was scattered onto Pluto-Charon/Arrokoth-crossing orbits.

To glean insights into what the crater size distributions might be like on Trojans, it is useful to consider the smallest impactors striking Charon and Arrokoth, two worlds from the primordial Kuiper belt where crater constraints exist via New Horizons observations. Despite the potential for numerous sub-km impactors to strike either body, both Charon and Arrokoth have a curious deficiency in the number of $D <$ 10 km diameter craters on their surfaces (Singer et al. 2019; Spencer et al. 2020; Robbins and Singer 2021). Charon craters follow a shallow cumulative power law slope of $q = -0.7 \pm 0.2$ for $1 < D < 10$ km craters (Singer et al. 2019; Robbins and Singer 2021), while Spencer et al. (2020) found Arrokoth craters had a slope of $q = -1.3$ for $0.2 < D < 1$ km craters. Note that a reassessment of both data sets by Morbidelli et al. (2021), which also considered the formation of a large 7 km crater on Arrokoth, argued that the power law slope of the crater size distribution was more likely to be $-1.5 < q < -1.2$, and that the projectile size distribution between 30 m and 1 km likely had $-1.2 < q < -1.0$.

Regardless, the shallow slope for a once sizeable population that is likely experienced collisional evolution is curious and intriguing, so its provenance may tell us about the physical properties of Trojans and primordial Kuiper belt objects. A similarly-shaped projectile SFD can be derived from small primary craters found on Europa, Ganymede, and Enceladus, worlds predominantly hit by comets from the scattered disk (Zahnle et al. 2003; Bierhaus et al. 2012). Morbidelli et al. (2021) argued this feature, mostly built during stage one and two collisions, was likely analogous to a similar shallow slope found in the asteroid belt. We will discuss this issue in Sec. 5, where we consider collisional evolution models for the Trojans.

# 5 Collisional Evolution Among the Trojans

## 5.1 Modeling the collisional evolution of the Trojans

Once the Trojans were captured in L4 and L5, they continued to undergo collisional evolution from impacting Trojans and, occasionally, from Hilda asteroids, Jupiter family comets, and nearly isotropic comets. We refer to these collisions as occurring in stage three. The nature of this evolution depends on several factors:

a. Their initial SFD at capture
b. Time of capture,
c. Collision probabilities and impact velocities of Trojans striking each other, and
d. Collisional disruption law followed by Trojans striking one another

While (a) is unknown, we can make some educated estimates as to what the SFD might have looked like and examine the amount of collisional grinding that potentially took place over the past ~4.5 Gyr. For example, the general similarity between the shapes of the Trojan and Kuiper belt SFDs for objects larger than several tens of km (see Sec. 2.3) might indicate the SFD for $D > 10$ km at the time of capture was not significantly different from the SFD today (Marschall et al. 2021). That would suggest the SFD for larger Trojans was set by some combination of planetesimal formation processes and early collisional grinding in a primordial Kuiper belt (Nesvorný et al. 2018; Nesvorný and Vokrouhlický 2019). We will discuss here how the SFD evolves in the case where the observed Trojan slope of $q = -2.1$ continues to small sizes.

For (b), we will assume that the giant planet instability takes place shortly after the dissipation of the solar nebula, as discussed in Sec. 4.2. For (c), input parameters were chosen similar to those in Nesvorný et al. (2018), namely that the intrinsic collision probability between Trojans within L4 or L5 are $P_i = 7 \times 10^{-18}$ km$^{-2}$ yr$^{-1}$ and that the impact velocity $V = 4.6$ km s$^{-1}$. Note that we purposely neglect the fact that the Hilda asteroids, Jupiter family comets, and nearly isotropic comets may also hit the Trojans, partly for simplicity but also because their influence on the overall collisional evolution of the Trojans is limited.

Concerning (d), the collisional strength of a body is commonly described by the critical impact energy $Q^*_D$, defined as the energy per unit target mass needed to disrupt and disperse 50% of the target (Benz and Asphaug 1999, see Davis et al. 2002). For the Trojans, the disruption law $Q^*_D$ is unknown and therefore must be treated as a free parameter.

Using the collisional model discussed in Marshall et al. (2021), we want to examine how the Trojans evolve from an initial SFD like the one discussed above when different $Q^*_D$ laws are applied. To interpret what is happening, it is useful to first consider what we know about collisional evolution in the main asteroid belt.

Consider that in the main belt, impacts on $D > 100$ km objects create numerous fragments, which themselves disrupt as part of a collisional cascade. Collisional modeling results indicate the easiest bodies to disrupt from an energy per mass perspective are $D \sim 200$ m (e.g., Bottke et al. 2020). This means $D < 200$ m objects, which are in the strength regime, grind themselves into a Dohnanyi-shaped SFD with a cumulative slope of $q \sim -2.6$ (Dohnanyi 1969; O'Brien and Greenberg 2003). This steep population of small objects goes on to disrupt bodies between 200 m and 2 km, creating a shallow slope of $q \sim -1$ between the inflection points. In turn, the paucity of bodies between 200 m and 2 km means that $D > 2$ km fragments produced by big collisions can build up over time, leading them to have a steep SFD. Overall, the models show the main belt SFD takes on a wavy shape.

Morbidelli et al. (2021) argued similar behavior would occur in the primordial Kuiper belt and its daughter populations, except there the weakest objects from an energy per mass perspective would be $D \sim$ 20 m rather than 200 m. This could potentially create the shallow slope found for $D < 10$ km craters on Charon and many giant planet satellites (see last section). The issue is whether these same trends hold for Trojans, where most of their bombardment for the last 4.5 Gyr has come from their fellow captured Trojans.

In **Fig. 3**, we set the weakest objects in the Trojan disruption scaling laws to be between $D = 20$ to 200 m. In each case, the bodies smaller than the weakest object develop a Dohnanyi-like slope of $q = -2.6$ (Dohnanyi 1969; O'Brien and Greenberg 2003). As expected, the sizes larger than the weakest body develop a shallow slope $q$ near -1. Overall, the behavior mimics collisional evolution in the main asteroid belt and probably the Kuiper belt/scattered disk (Bottke et al. 2005; 2015; 2020).

PLACE FIGURE 3 HERE

Though the shallow slope at intermediate sizes occurs in all cases investigated, the details of where the shallow slope begins and ends depends strongly on $Q^*_D$. Determining where this collisional break occurs will critically inform the nature of the Trojan disruption law. We predict the crater SFDs found on Lucy's Trojan targets may be the key data needed to understand the collisional evolution of the Trojans and other small outer solar system bodies.

### 5.2 The Eurybates family in the Trojan population

Another important feature of collisional evolution in a population is the formation of collisional families through large catastrophic impacts. Within the Trojans there is only one large collisional family: Eurybates, with (3548) Eurybates, a C-type Trojan, as its largest member family (Brož and Rozehnal 2011). The Eurybates family is well within the L4 stability region (Robutel and Gabern 2006). In inclination, the family is well defined, with most members within half a degree of 7.5° (Brož and Rozehnal 2011) (**Fig. 1**). From a spectral signature standpoint, the Eurybates family is also bluer than the overall L4 population. Others within the Trojans include the Hektor and Arkesilaos families (e.g. Rozehnal et al. 2016, Holt et al. 2020), but they will not be discussed further here.

The largest ten members of the Eurybates family, with diameters between 20 and 65 km, have a SFD whose power law slope is consistent with the overall slope of the Trojan population. At $D < 20$ km, the family's SFD is considerably steeper than the background population, a result indicative of a collisional family. The shape of the SFD, being diagnostic of the type of impact that formed the family, suggests that a 100 km parent body was hit by a ~35 km impactor (Marschall et al. 2022). Further, the SFD indicates that the parent body had a rather weak collisional strength consistent with impact simulations of large rubble piles (Benavidez et al. 2012).

(3548) Eurybates has a km-sized satellite called Queta (Noll et al 2020). It was likely created by disruptive collision that formed the family. Queta's existence makes it possible to estimate the family's approximate age, though it depends on the nature of the disruption law employed. Using the Benz and Asphaug (1999) disruption law for ice, which yields the Trojan SFD shown by the purple curves in **Fig. 2**, the model age for the Eurybates family is $< 3.7$ Gy (Marschall et al. 2021). Previous family ages have resulted in a range from 1 Gyr to 4 Gyr (Brož and Rozehnal 2011; Rozehnal et al. 2016; Milani et al. 2017; Holt et al. 2020).

Though collisionally-derived satellites are a common byproduct of catastrophic disruption events, numerical models suggest they should have large eccentricities (Durda et al. 2004). Queta, with an eccentricity of only 0.125 (Brown et al. 2021), is not a typical case in that respect, though the Kozai resonance could produce eccentricity/inclination oscillations that raise the eccentricity to 0.4 (see chapter by Levison et al., this volume). With that said, its low eccentricity could also be telling us that better collisional models are needed. Regardless, Queta's formation and survival requires further explanation.

**5.3 Colors of Trojans**

Additional insights on Trojan evolution come from their colors and spectra. As discussed above, all Trojans tend to have low albedos. Objects with very red featureless slopes in visible and near infrared wavelengths are classified as D-type asteroids, those with modestly red slopes are P-types asteroids, and those with flat or nearly horizontal slopes are C-type asteroids (e.g., DeMeo et al. 2015). Obtaining detailed spectroscopic observations for objects more distant than Trojans can be difficult, though, so taxonomic information is limited for those worlds. For this reason, most observers have concentrated on colors, which provide a sense of the spectral slope of outer solar system worlds. Different groups have done this in different ways.

For example, one way is to use spectral slope to make color distinctions, with less red (LR) objects defined as those with spectra slopes near $5 \times 10^{-5}$ Å$^{-1}$, red (R) objects having spectral slopes near $10 \times 10^{-5}$ Å$^{-1}$, and very red (VR) objects having values that are considerably larger, roughly $> 20$ to $25 \times 10^{-5}$ Å$^{-1}$ (Wong et al. 2014; Wong and Brown 2015; 2016; 2017a). Other groups use the *ugriz* magnitude system to make these color assessments (see Nesvorný et al 2020 and references therein) or specific observation filters (Jewitt 2002; Jewitt 2018). For this discussion, we will lean on the LR, R, and VR definitions, with the boundaries between LR, R, and VR being modestly flexible. From this background, we can define what we know of various outer solar system populations.

The Trojans and Hildas have bimodal distribution of LR and R objects, and both resemble each other (Wong and Brown 2016; 2017a,b). The similarity is consistent with the idea that they have a common origin and source region. No VR objects have been identified in either population to date, an issue we will discuss below. The large asteroid families in the Trojans (i.e., Eurybates) and Hildas (i.e., Hilda, Schubert) are made up of LR objects. This suggests that collisional disruption exposes interior materials of the parent bodies which are LR.

As we move further out in the solar system, the color distributions become more complex and diverse, as shown in **Fig. 4**, which has been adapted from Bolin et al. (2023) (see also Jewitt 2018).

- **Neptune Trojans**: These bodies have a LR/R color distribution similar to the Trojans, but include some VR objects.
- **Centaurs**: Nearly 20% of the population is comprised of VR bodies.
- **Scattered Disk**: It is the source of the Centaurs, and has color trends similar to that population.
- **Plutinos**: A 50:50 mix of R and VR objects.
- **Hot Classical**: This population has a comparable fraction of VR bodies to the Centaurs and scattered disk.
- **Cold Classicals**: They are dominated by VR objects, though a few objects are R and trend toward LR.

PLACE FIGURE 4 HERE

For this exercise, we will assume that all of the above color distributions can be explained within the context of the giant planet instability model, which provides the best available explanation of the dynamical properties of the above populations. This leads to different ways to interpret these color trends.

We also must point out a caveat, namely that the majority of the Trojans and Centaurs characterized above are mainly bodies with diameters smaller than a few tens of km, while the characterized objects in more distant populations like the scattered disk, Plutinos, and various Kuiper belt populations are dominated by $D > 100$ km bodies. This size difference may mean that some comparisons below are potentially inappropriate.

Our first comment on **Fig. 4** is that Jupiter and Neptune Trojans (and the Hildas, which are not plotted in Jewitt 2018 or Bolin et al. 2024) have close to the same color distribution. Dynamically, the bodies captured in the Neptune Trojans stayed far for the Sun, so that suggests that outgassing played a minimal role in producing the observed colors of either population, and that the LR objects in the Neptune Trojans were probably not produced this way.

Our second comment is that there are no VR colors in the Jupiter Trojans, but a small fraction exists in the Neptune Trojans. These data provide an update to Jewitt (2018); at that time of that publication, no VR Neptune Trojans were known. Objects with VR colors are more plentiful in the Centaurs, scattered disk, and other populations.

The dichotomy between the Jupiter and Neptune Trojans for VR objects may hold interesting information. Nesvorný et al. (2020) showed that most of Neptune's Trojans come from 25-30 au within the primordial Kuiper belt, so placing the transition near 30 au could potentially explain why the Jupiter Trojans lack VR objects (Bolin et al. 2023). This would also imply that most VR objects are from a low mass extension of the primordial Kuiper belt between ~30-40 au, namely the zone that lies between Neptune's current orbit at 30 au and the current location of the cold classical Kuiper belt. Nesvorný et al. (2020) also point out that having the VR objects from that zone would explain why they tend to have lower inclinations; objects destabilized by Neptune in this region do not get as excited as those between 20-30 au.

An alternative way to explain the dichotomy is to argue it is a surface veneer that goes away close to the Sun. Wong and Brown (2016; 2017a) have argued that VR colors may be produced by the sublimation of hydrogen sulfide ice at distances greater than ~15-20 au, with VR bodies then scattered to many small body reservoirs by Neptune's outward migration. The absence of VR bodies in Jupiter's Trojans and Hildas might represent the loss of surface volatiles and/or organics via solar heating close to the Sun, with the VR objects transformed into R objects. Such a transition has been observed by Jewitt (2015), with VR Centaurs gradually losing this color as they pass within 10 au. Nesvorný et al. (2020) suggested sublimation-driven surface depletion of some organic molecules, such as $NH_3$, between 30-40 au might explain the R-VR transition (Brown et al. 2011).

Our third and last comment is that the role of collisional evolution in explaining these color trends has yet to be fully explored. Asteroid families in the Trojan and Hilda populations demonstrate that one way of producing LR objects is by collisional disruption (e.g., the Trojan target Eurybates). It is unclear what happens if collisions occur in other populations far from the Sun.

## 6 Conclusions

As this discussion shows, the origin and evolution of the Trojans are linked to the origin and evolution of planetesimals in the primordial Kuiper belt and the nature of giant planet migration. Yet, it is possible that Lucy's flybys past these diverse worlds will revolutionize our ideas about these bodies to such a degree that some of the suggested models discussed above will be overturned. The work done by this initial reconnaissance will lead to follow-up studies, perhaps involving landings and sample return missions to P- and D-types that are close to the Earth than the Trojans. This will potentially allow us to test the hypothesis that all such bodies derive from the primordial Kuiper belt.

**Acknowledgements**

We thank the members of the Lucy team for their helpful comments in making this a better review article. We also thank our anonymous referee for their valuable comments that made this a better paper. The work in this paper was supported by NASA's Lucy mission through contract NNM16AA08C. The work of David Vokrouhlický was partially funded 2601 by grant 21-11058S from the Czech Science Foundation.

Sekine, Y., Genda, H., Kamata, S., Funatsu, T. 2017. The Charon-forming giant impact as a source of Pluto's dark equatorial regions. Nature Astronomy 1, 0031.

Shoemaker, E. M., C. S. Shoemaker, and R. F. Wolfe. 1989. Trojan asteroids: Populations, dynamical structure and origin of the L4 and L5 swarms. In Asteroids II (R. P. Binzel, T. Gehrels, and M. S. Matthews, Eds.) U. Arizona Press, Tucson, 921–948.

Singer, K. N., W. B. McKinnon, S. Greenstreet, B. Gladman, E. B. Bierhaus, S. A. Stern, A. H. Parker, et al., 2019. Impact craters on Pluto and Charon indicate a deficit of small Kuiper belt objects. Science 363: 955-959.

Spencer J. R., S. A. Stern, J. M. Moore, H. A. Weaver, K. N. Singer, C. B. Olkin, A. J. Verbiscer, W. B. McKinnon, et al. 2020. The geology and geophysics of Kuiper Belt object (486958) Arrokoth. Science 367: eaay3999.

Stern, S. A., O. L. White, W. M. Grundy, B. A. Keeney, J. D. Hofgartner, D. Nesvorný, W. B. McKinnon, D. C. Richardson, J. C. Marohnic, A. J. Verbiscer, S. D. Benecchi, P. M. Schenk, J. M. Moore, New Horizons Geology, and Geophysics Investigation Team. 2023. The Properties and Origin of Kuiper Belt Object Arrokoth's Large Mounds. The Planetary Science Journal 4: 176.

Szabó, G. M., Ž. Ivezić, M. Jurić, and R. Lupton. 2007. The properties of Jovian Trojan asteroids listed in SDSS Moving Object Catalogue 3. Monthly Notices of the Royal Astronomical Society 377: 1393-1406.

Szabó, G. M., A. Pál, C. Kiss, L. L. Kiss, L. Molnár, O. Hanyecz, E. Plachy, K. Sárneczky, and R. Szabó. 2017. The heart of the swarm: K2 photometry and rotational characteristics of 56 Jovian Trojan asteroids. Astronomy and Astrophysics 599: A44.

Tsiganis, K., R. Dvorak, and E. Pilat-Lohinger. 2000. Thersites: a "jumping" Trojan? Astronomy and Astrophysics 354: 1091-1100.

Tsiganis, K., R. Gomes, A. Morbidelli, and H. F. Levison. 2005. Origin of the orbital architecture of the giant planets of the Solar System. Nature 435: 459-461.

Uehata, K., Terai, T., Ohtsuki, K., Yoshida, F. 2022. Size Distribution of Small Jupiter Trojans in the L5 Swarm. The Astronomical Journal 163, 213. doi:10.3847/1538-3881/ac5b6d

Vokrouhlický, D., W. F. Bottke, and D. Nesvorný. 2016. Capture of Trans-Neptunian Planetesimals in the Main Asteroid Belt. The Astronomical Journal 152: 39. DOI: 10.3847/0004-6256/152/2/39.

Vokrouhlický, D., D. Nesvorný, and L. Dones. 2019. Origin and Evolution of Long-period Comets. Astron. J. 157: 181.

Walsh, K. J., A. Morbidelli, S. N. Raymond, D. P. O'Brien, and A. M. Mandell. 2011. A low mass for Mars from Jupiter's early gas-driven migration. Nature 475: 206-209. DOI: 10.1038/nature10201.

Wong, I. and M. E. Brown. 2015. The Color-Magnitude Distribution of Small Jupiter Trojans. The Astronomical Journal 150: 174.

Wong, I. and M. E. Brown. 2016. A Hypothesis for the Color Bimodality of Jupiter Trojans. The Astronomical Journal 152: 90.

Wong, I. and M. E. Brown. 2017a. The Bimodal Color Distribution of Small Kuiper Belt Objects. The Astronomical Journal 153: 145.
25

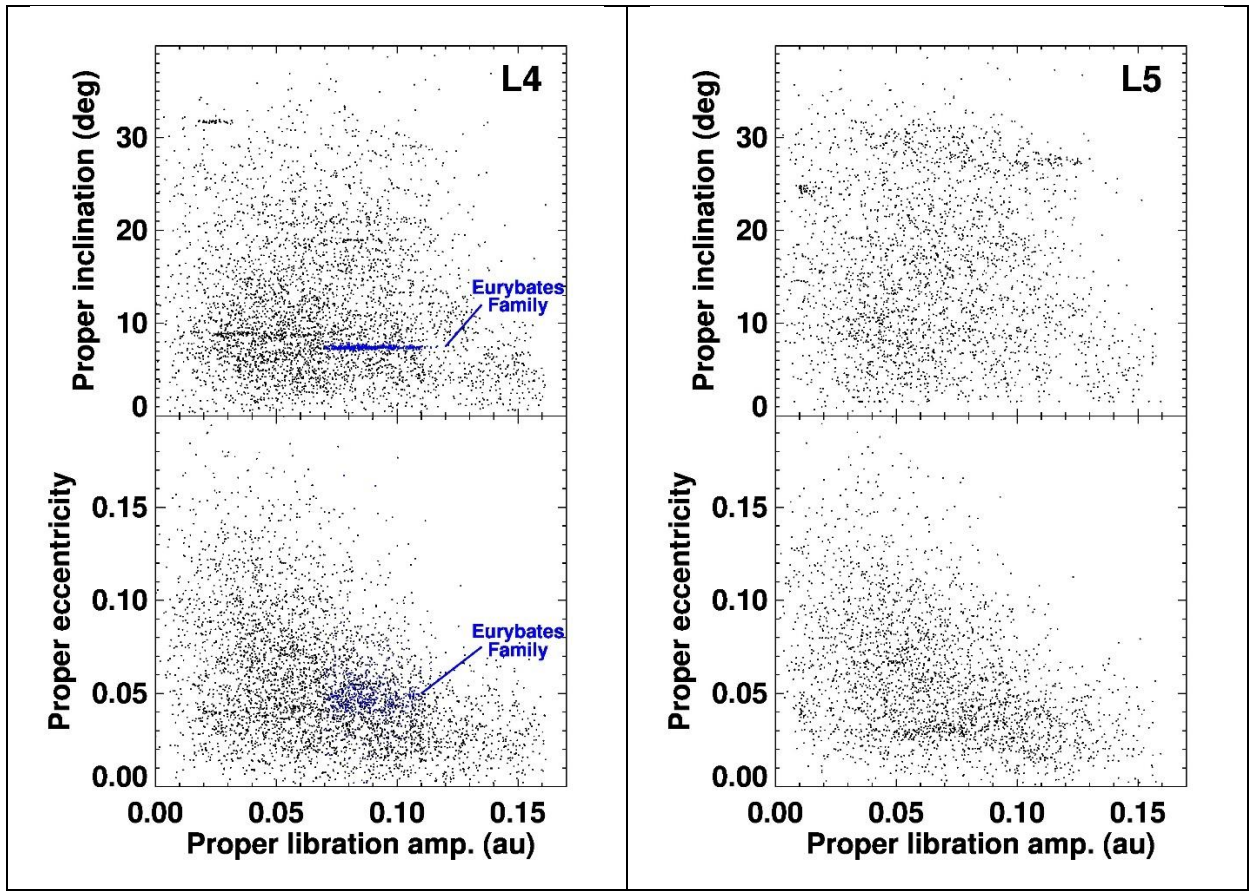

**Figure 1**: Proper orbital elements of Jovian Trojans. The left two panels show the orbits of 4607 objects at the L4 libration zone, while the right two panels show the orbits of 2721 objects at the L5 libration zone (see Holt et al. 2020 for discussion of orbital elements). The Eurybates family is shown in L4 as the blue dots. The proper orbital elements, namely libration amplitude, eccentricity, inclination, of the Trojans were calculated using the methods of Milani (1993). The libration amplitude-eccentricity plots highlight the stable zone for Trojans; as objects approach the top-right corner, they reach unstable orbits and can escape.

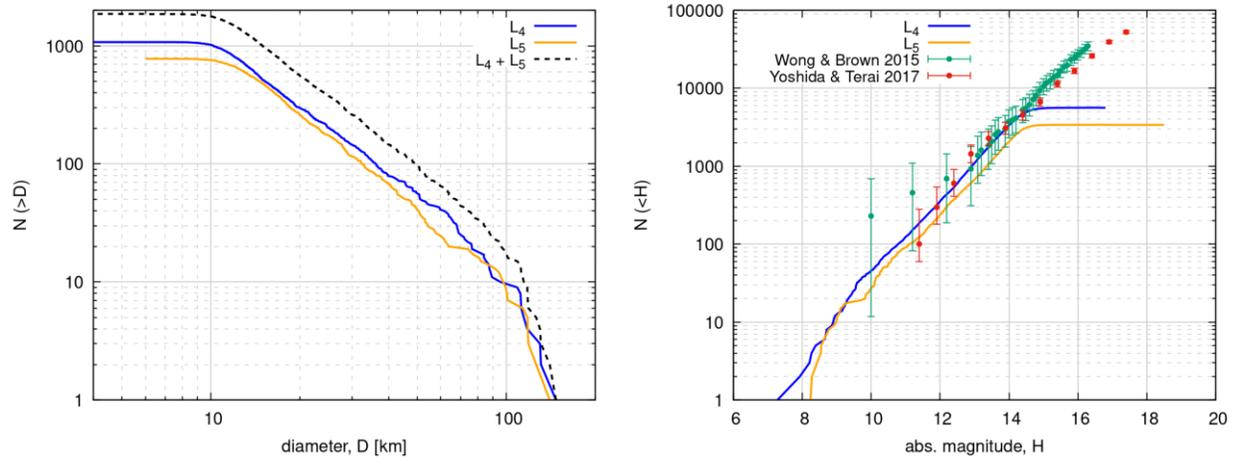

**Figure 2:** The Trojan cumulative size frequency distribution (SFD) as a function of WISE diameter (left panel) and absolute magnitude (right panel) are shown for the L4 (orange) and L5 (blue) swarm (e.g., Mainzer et al. 2019; Marschall et al. 2022). The WISE SFDs are for those Trojans observed by WISE, so the curves should be interpreted in terms of shape and not absolute number. In addition, the right plot includes data from the L4 pencil beam studies of Wong and Brown (2015) and Yoshida and Terai (2017).

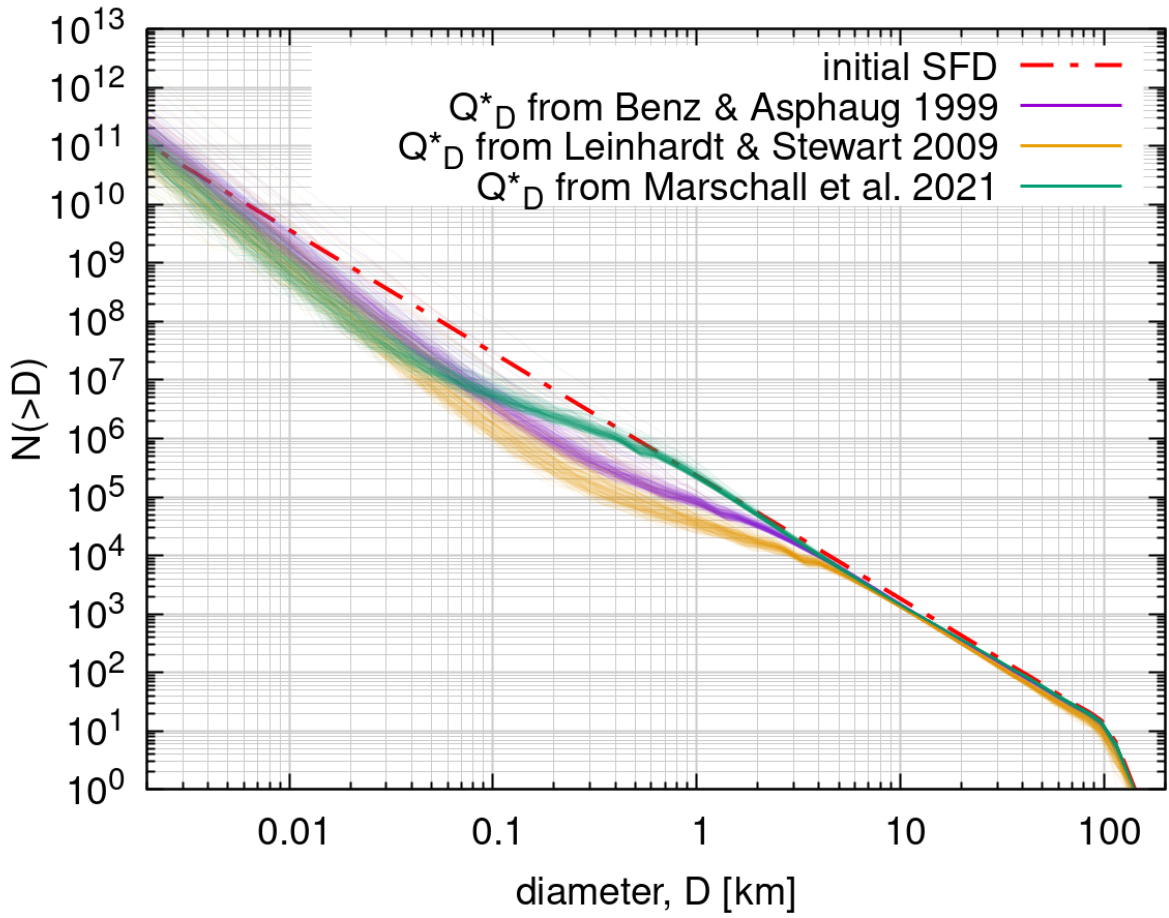

**Figure 3**: Trojans size distribution after 4.5 Gy of collisional evolution for three different Q*D. The initial SFD is shown in the dashed red line. Adapted from Marschall et al. (2022).

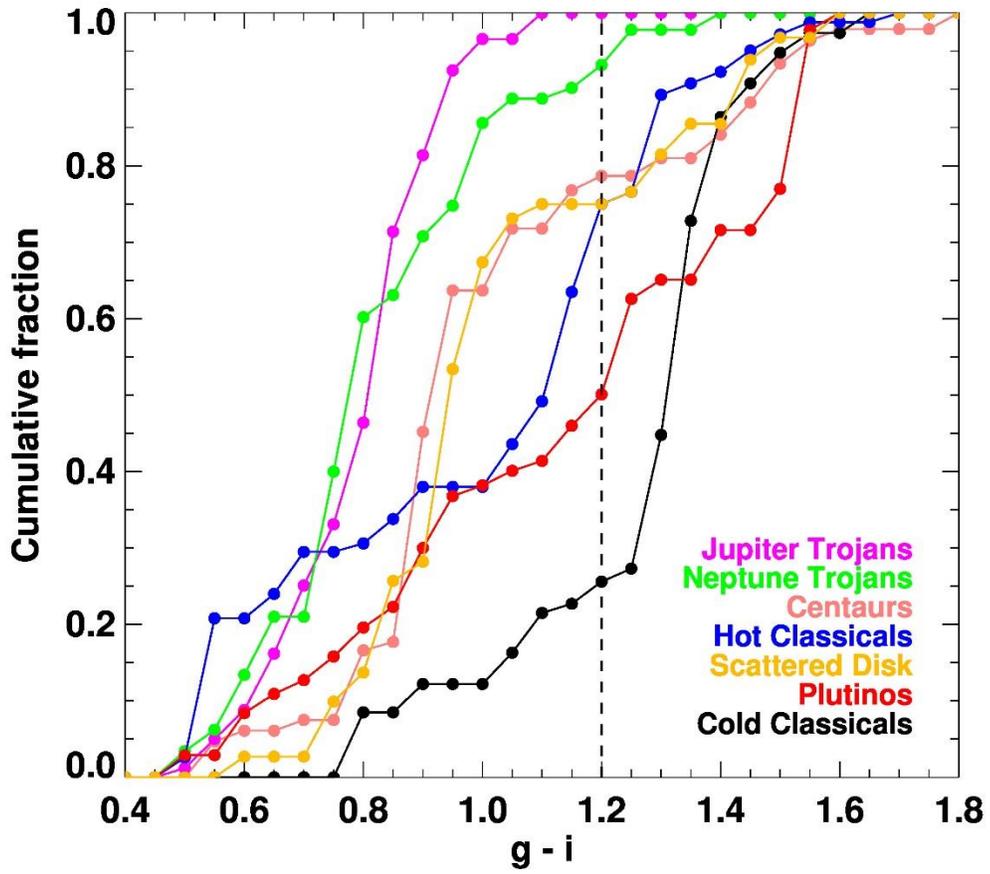

**Figure 4.** Cumulative distributions of the g - i color index for many different outer solar system populations (see Sec 5.3). Ultrared objects are assumed to be to the right of the dashed vertical line. Adapted from Bolin et al. (2023).